\documentstyle[twocolumn,aps,prl,epsfig]{revtex}
\newcommand{\beq}{\begin{equation}}
\newcommand{\eeq}{\end{equation}}
\newcommand{\eq}[1]{Eq. (\ref{#1})}

\def\prep#1#2#3{Phys.~Rep. ~{\bf #1},\ #2\ (#3)}

\def\prl#1#2#3{Phys.~Rev.~Lett.~{\bf #1},\ #2\ (#3)}

\def\pra#1#2#3{Phys.~Rev.~A~{\bf #1},\ #2\ (#3)}

\def\ao{{\hat a}_1}
\def\aod{{\hat a}_1^\dagger}
\def\at{{\hat a}_2}
\def\atd{{\hat a}_2^\dagger}
\def\aj{{\hat a}_j}

\def\lx{{\hat L}_x}
\def\ly{{\hat L}_y}

\def\lz{{\hat L}_z}
\def\li{{\hat L}_i}
\def\lj{{\hat L}_j}
\def\lk{{\hat L}_k}

\def\u{S_x}
\def\v{S_y}
\def\w{S_z}
\def\s{{\bf{\vec s}}}
\def\udot{{\dot S}_x}
\def\vdot{{\dot S}_y}
\def\wdot{{\dot S}_z}
\def\dxz{\Delta_{xz}}
\def\dyz{\Delta_{yz}}
\def\dxy{\Delta_{xy}}
\def\dzz{\Delta_{zz}}
\def\dyy{\Delta_{yy}}
\def\dxx{\Delta_{xx}}
\def\dij{\Delta_{ij}}
\def\dxzdot{{\dot\Delta}_{xz}}
\def\dyzdot{{\dot\Delta}_{yz}}
\def\dxydot{{\dot\Delta}_{xy}}
\def\dzzdot{{\dot\Delta}_{zz}}
\def\dyydot{{\dot\Delta}_{yy}}
\def\dxxdot{{\dot\Delta}_{xx}}
\def\dli{{\hat{\delta L}}_i}
\def\dx{\delta_x}
\def\dy{\delta_y}
\def\dz{\delta_z}
\def\di{\delta_i}
\def\dj{\delta_j}
\def\dxdot{{\dot\delta}_x}
\def\dydot{{\dot\delta}_y}
\def\dzdot{{\dot\delta}_z}
\def\k{\eta}
\def\ddt{\frac{d}{dt}}

\begin{document}

\title{Condensates beyond mean field theory: quantum backreaction as 
decoherence}
\author{A. Vardi and J.R. Anglin}
\address{ITAMP, Harvard-Smithsonian Center for Astrophysics
60 Garden Street, Cambridge MA 02138}
\date{\today}
\maketitle

\begin{abstract}
{\bf Abstract}~~~
We propose an experiment to measure the slow $\log (N)$ convergence to
mean-field theory (MFT) around a dynamical instability. Using a density 
matrix formalism, we derive equations of motion which go beyond MFT and
provide accurate predictions for the quantum break-time. The leading 
quantum corrections appear as decoherence of the reduced single-particle 
quantum state.
\end{abstract}
\maketitle

A Bose-Einstein condensate is described in mean field theory (MFT) by a c-number 
macroscopic wave function, obeying the Gross-Pitaevskii non-linear Schr\"odinger 
equation.  MFT is closely analogous to the semiclassical approximation of 
single-particle quantum mechanics, with the inverse square root of the number 
$N$ of particles in the condensate playing the role of $\hbar$ as a perturbative 
parameter.  Since in current experimental condensates $N$ is indeed large, it is 
generally difficult to see qualitatively significant quantum corrections to MFT. 
 In the vicinity of a dynamical instability in MFT, however, quantum corrections 
appear on timescales that grow only logarithmically with $N$.  In this paper we 
propose an experiment to detect such quantum corrections, and present a simple 
theory to predict them.  We show that, as the Gross-Pitaevskii classical limit 
of a condensate resembles single-particle quantum mechanics, so the leading 
quantum corrections appear in the single-particle picture as decoherence.

We will consider a condensate in which particles can only effectively populate 
two second-quantized modes. This model can be realized with a condensate in a 
double well potential\cite{javanainen,walls,Janne,smerzi,leggett,lewenstein}, or 
with an effectively two-component spinor condensate\cite{cornell1,cornell2} 
whose internal state remains uniform in space.  Such uniformity can be ensured, 
to a good approximation, by confining a very cold condensate within a size much 
smaller than $[n|\sqrt{a_{11}a_{22}}-a_{12}|]^{-1/2}$, where $n$ is the mean 
total density and $a_{ij}$ is the s-wave scattering length between atoms in 
internal states $i$ and $j$.  The kinetic energy of spin non-uniformity then 
ensures that 
the spatial state of the condensate will adiabatically follow its internal 
state, which will evolve on slower time scales.  Dynamical instabilities to 
phase separation are also frustrated in this regime, which since for available 
alkali gases all $a_{ij}$ differ only by a few percent, could be reached with 
small condensates ($N\leq10^4$) in weak, nearly spherical traps ($\omega\leq 
100$ Hz).  Stronger or less isotropic traps reach the two-mode regime at smaller 
$N$.

In the double well realization, the nonlinear interaction may be taken to affect 
only atoms within the same well.  In this case single-particle tunneling 
provides a linear coupling between the two modes, which can in principle be 
tuned over a wide range of strengths.
Two internal states may be coupled by a near-resonant radiation field 
\cite{cornell3,stenholm}.  If collisions do not change spin states, there is 
also a simple nonlinear interaction in the internal realization.  In either case 
the total number operator commutes with the Hamiltonian, and may be replaced 
with the c-number $N$.  Discarding c-number terms, we may therefore write the 
two-mode Hamiltonian
\beq
H=-\frac{\omega}{2}\left(\aod\at+\atd\ao\right)+
\frac{\k}{2}\left[(\aod)\ao-(\atd)\at\right]^2~,
\label{ham}
\eeq 
where $\omega$ is the coupling strength between the two condensate 
modes, $\k$ is the two-body interaction strength, 
and $\ao,\aod,\at,\atd$ are particle annihilation and creation operators 
for the two modes.  We will take $\k$ and $\omega$ to be positive, since the 
relative phase between the two modes may be re-defined arbitrarily, and since 
without dissipation the overall sign of $H$ is insignificant.

Instead of considering the evolution of $\aj$ and its expectation value in a 
symmetry-breaking ansatz, we will examine the evolution of the directly 
observable quantities $\hat{a}_i^\dagger\hat{a}_j$, whose expectation values 
define the reduced single particle density matrix (SPDM) $R_{ij}\equiv 
\langle\hat{a}_i^\dagger\hat{a}_j\rangle/N$. It is convenient to introduce the 
Bloch representation, by defining the angular momentum operators,
$$
\lx\equiv\frac{\aod\at+\atd\ao}{2}~,~\ly\equiv\frac{\aod\at-\atd\ao}{2i}
$$
\beq
\lz=\frac{\aod\ao-\atd\at}{2}~.
\label{lxyz}
\eeq
The Hamiltonian \eq{ham} then assumes the form,
\beq
H=-\omega\lx+\frac{\k}{2}\left(\frac{\hat{N}}{4}^2+\lz^2\right)\;,
\label{hamiltonian}
\eeq
and the Heisenberg equations of motion for the three angular momentum 
operators of \eq{lxyz} read
\begin{eqnarray}\label{Ldot}
\ddt\lx=-i[\lx,H]&=&-\frac{\k}{2}(\ly\lz+\lz\ly)~,\nonumber\\
\ddt\ly=-i[\ly,H]&=&+\omega\lz+\frac{\k}{2}(\lx\lz+\lz\lx)~,\\
\ddt\lz=-i[\lz,H]&=&-\omega\ly~.\nonumber
\end{eqnarray}

The mean-field equations for the SPDM in the two-mode model 
may be obtained, without invoking U(1) symmetry breaking, by approximating 
second-order 
expectation values $\langle\li\lj\rangle$ as products of the first 
order expectation values $\langle\li\rangle$ and $\langle\lj\rangle$:
\beq
\langle\li\lj\rangle\approx \langle\li\rangle\langle\lj\rangle~.
\label{mfa}
\eeq
Defining the single-particle Bloch vector ${\bf{\vec s}}=(\u,\v,\w)
=(\frac{2\langle\lx\rangle}{N},\frac{2\langle\ly\rangle}{N}
,\frac{2\langle\lz\rangle}{N}),~\kappa=\k N/2$ 
and using \eq{mfa}, we obtain the nonlinear Bloch equations
\begin{eqnarray}\label{nlbloch} 
\udot&=&-\kappa \w\v ~,\nonumber\\
\vdot&=&\omega \w+\kappa \w\u ~,\\
\wdot&=&-\omega \v ~.\nonumber
\end{eqnarray}
Mean field trajectories ${\bf{\vec s}}(t)$ at four different 
$\kappa/\omega$ ratios are plotted in Fig. 1. 
The norm of ${\bf{\vec s}}$ is conserved in MFT, and so for an initially pure 
SPDM, (\ref{nlbloch}) are equivalent to the two-mode Gross-Pitaevskii equation 
\cite{smerzi}. 
\begin{figure}
\begin{center}
\epsfig{file=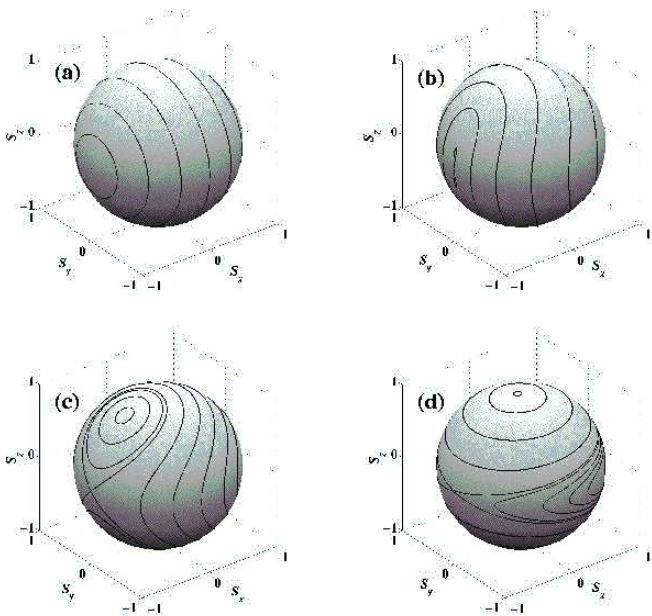,width=\columnwidth}
\end{center}
\caption{ Mean-field trajectories at (a)$\kappa=0$, (b)$\kappa=1.02\omega$, 
(c)$\kappa=2\omega$, and (d)$\kappa=20\omega$. }
\label{f1 }
\end{figure}
The nonlinear Bloch equations (\ref{nlbloch}) depict a competition between 
linear Rabi oscillations in the $\v\w$-plane and nonlinear 
oscillations in the $\u\v$-plane. For a noninteracting condensate
(Fig. 1a) the trajectories on the Bloch sphere are circles about the $\u$ axis, 
corresponding to harmonic Rabi oscillations. As $\kappa$ increases 
the oscillations become more anharmonic until, above the critical value 
$\kappa=\omega$ (Fig. 1b), the stationary point ${\bf{\vec s}}=(-1,0,0)$ 
becomes dynamically unstable, and macroscopic self-trapping can occur 
(oscillations with a non-vanishing time averaged 
population imbalance  $\langle \w\rangle_t\neq 0$) \cite{smerzi}.  In the 
vicinity of the dynamically unstable point, MFT will break down on a time scale 
only logarithmic in $N$, and so an improved theory is desirable.

If we assume that the condensate remains mildly fragmented, so that the two
eigenvalues of $R_{ij}$ are $f$ and $1-f$ for small $f$,
we can take $\li=L_i+\dli$, where the c-number $L_i$ is ${\cal O}(N)$ and, 
throughout the Hilbert subspace through which the system will evolve, all matrix 
elements of $\hat{\delta L}_i$ are no greater than ${\cal O}(N\sqrt{f})$.
The second order moments
\beq\label{dij}
\dij=4N^{-2}\left(\langle\li\lj+\lj\li\rangle
-2\langle\li\rangle\langle\lj\rangle\right)~,
\eeq
will then be of order $f$.  We can retain these, and so improve on MFT,
if we truncate the BBGKY hierarchy of expectation value equations of motion 
at one level deeper: we eliminate the approximation (\ref{mfa}), and instead 
impose
$$
\langle\li\lj\lk\rangle\approx\langle\li\lj\rangle\langle\lk\rangle+
\langle\li\rangle\langle\lj\lk\rangle+\langle\li\lk\rangle\langle\lj\rangle
$$
\beq\label{trunc2}
-2\langle\li\rangle\langle\lj\rangle\langle\lk\rangle\;.
\eeq
This approximation is accurate to within a factor $1+{\cal O}(f^{3/2})$, better 
than (\ref{mfa}) by one factor of $f^{1/2}$.  Successively deeper truncations of 
the hierarchy yield systematically better approximations as long as $f$ is 
small.  

Applying (\ref{dij}) and (\ref{trunc2}) to (\ref{Ldot}), we obtain the following 
set of equations, in which the mean field Bloch vector drives the fluctuations 
$\dij$, and is in turn subject to backreaction from them:
\begin{eqnarray}\label{bgb}
\udot&=&-\kappa\w\v-\frac{\kappa}{2}\dyz\nonumber\\
\vdot&=&\omega\w+\kappa\w\u+\frac{\kappa}{2}\dxz\nonumber\\
\wdot&=&-\omega\v\nonumber\\
\dxzdot&=&-\omega\dxy-\kappa\w\dyz-\kappa\v\dzz\nonumber\\
\dyzdot&=&\omega(\dzz-\dyy)+\kappa\w\dxz+\kappa\u\dzz\\
\dxydot&=&(\omega+\kappa\u)\dxz-\kappa\v\dyz+\kappa\w(\dxx-\dyy)\nonumber\\
\dxxdot&=&-2\kappa\v\dxz-2\kappa\w\dxy\nonumber\\
\dyydot&=&2(\omega+\kappa\u)\dyz+2\kappa\w\dxy\nonumber\\
\dzzdot&=&-2\omega\dyz~.\nonumber
\end{eqnarray}
In what follows, we will refer to (\ref{bgb}) as evolution under 
`Bogoliubov backreaction' (BBR).

We note that (\ref{bgb}) are actually identical to the equations of motion one 
would obtain, for the same quantities, using the Hartree-Fock-Bogoliubov 
Gaussian ansatz.   
And if the second-order moments $\dij$ may initially be factorized as 
$\dij=\di\dj$ ($i,j=x,y,z$), then the factorization persists 
and the time evolution 
of $\dx,\dy$, and $\dz$ is equivalent to that of perturbations of the 
mean-field equations (\ref{nlbloch}):
\begin{eqnarray}\label{bog2}
\dxdot&=&-\kappa(\w\dy+\v\dz)~,\nonumber\\
\dydot&=&\omega\dz+\kappa(\w\dx+\u\dz)~,\\
\dzdot&=&-\omega\dy~.\nonumber
\end{eqnarray}
Thus our equations for $\dij$ are in a sense equivalent to the usual Bogoliubov 
equations.  The quantitative advantage of our approach therefore lies entirely 
in the wider range of initial conditions that it admits, which may more 
accurately represent the exact initial conditions.  For instance, a Gaussian 
approximation will have $\dxx={\cal O}(1)$ in the ground state, where in fact 
$\dxx = {\cal O}(N^{-1})$.  This leads to an error of order $N^{-1/2}$ in the
Josephson frequency computed by linearizing (\ref{bgb}) around the ground state,
which is problematic because one would hope that the Gaussian backreaction 
result would be accurate at this order.  Our SPDM approach avoids this problem, 
which is presumably the two-mode analogue of the Hartree-Fock-Bogoliubov 
spectral gap\cite{Griffin}.

For finite motion of the Bloch vector, our formalism offers an efficient method 
to depict the back-reaction of the Bogoliubov 
equations 
on the mean field equations via the 
coupling terms $-\kappa\dyz/2$ and $\kappa\dxz/2$ in (\ref{bgb}). Because 
in general $\dyz(t)\neq\dxz(t)$, this back-reaction has the
effect of breaking the unitarity of the mean-field dynamics.
Consequently, the BBR trajectories are
no longer confined to the surface of the Bloch sphere, but 
penetrate to the interior (representing mixed-state $R_{ij}$, 
with two non-zero eigenvalues).  Thus although decoherence is generally
considered as suppressing quantum effects, decoherence of the
single particle quantum state of a condensate is itself the leading quantum 
correction (due to interparticle entanglement) to the effectively 
classical MFT. 
\begin{figure}
\begin{center}
\epsfig{file=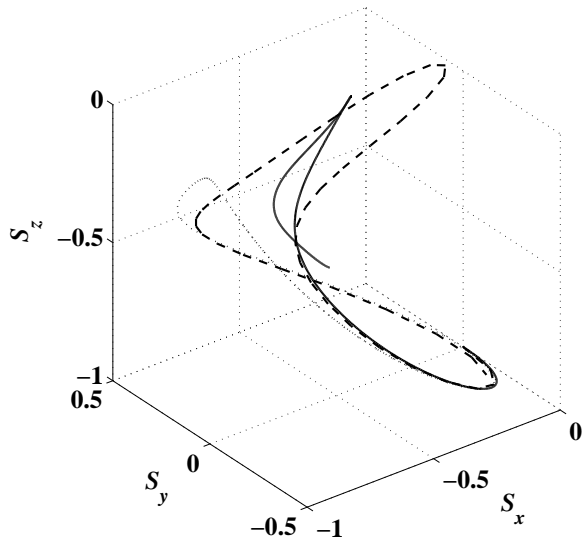,width=0.9\columnwidth}
\end{center}
\caption{Mean field ($\cdot\cdot\cdot$), Bogoliubov back-reaction ($---$) and 
exact 50 particles (-----) trajectories starting with all particles in one
mode, at $\kappa=2\omega$.}
\label{f3}
\end{figure}
In order to confirm this decoherence, and demonstrate how the BBR 
equations (\ref{bgb}) improve on the mean-field equations 
(\ref{nlbloch}), we compare the trajectories obtained by these 
two formalisms to exact quantum trajectories, obtained by 
fixing the total number of particles $N=50$ and solving \eq{Ldot} 
numerically, using a Runge-Kutta algorithm.  The results for 
an initial state where all particles are in one of the modes 
(corresponding to the initial conditions $\u=\v=0,\w=-1,
\dxx=-\dyy=2/N,\dxy=\dxz=\dyz=\dzz=0$)
and $\kappa=2\omega$ are shown in Fig. 2.  The 
MFT trajectory passes through the dynamically unstable point 
${\bf{\vec s}}=(-1,0,0)$. Consequently, the quantum trajectory 
sharply breaks away from the MFT trajectory as it approaches 
this point, entering the Bloch sphere interior.  
While still periodic on a much shorter time scale than the exact evolution, 
the BBR
evolution (dashed curve) provides an excellent prediction of the time at which 
the break from MFT takes place (the 'quantum break time').   
\begin{figure}
\begin{center}
\epsfig{file=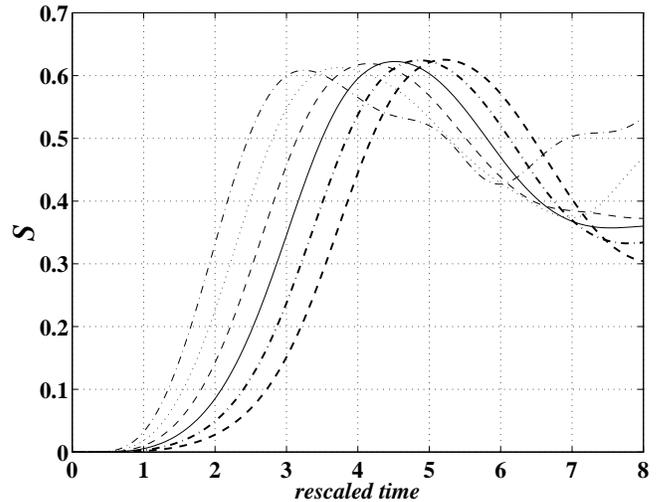,width=\columnwidth}
\end{center}
\caption{Growth of the von Neumann entropy $S$ of the quantum reduced 
single-particle density operator, at $\kappa=2\omega$, for N=10 ($-\cdot-$), 
20 ($\cdot\cdot\cdot$), 40 ($---$), 80 (------), 
160 ({\bf - . - . -}), and 320 ({\bf - - - -}) 
particles. Initial conditions are the same as in Fig. 2. }
\label{f4 }
\end{figure}
The quantum break time near a dynamical instability is expected 
to grow logarithmically with N. In Fig.~3 we plot the 
von Neumann entropy 
\beq
S=-\frac{1}{2}\ln\left[\frac{(1+|\s|)^{(1+|\s|)}(1-|\s|)^{(1-|\s|)}}{4}\right]
\eeq 
of the exact reduced single-particle 
density operator, as a function of the rescaled time $\omega t$ 
with N=10,20,40,80,160, and 320 particles, for the same initial
conditions as in Fig.~2. Since the MFT entropy is always zero, 
$S$ serves as a measure of convergence. The quantum break time 
is clearly evident, and indeed increases as $\log(N)$. The 
single-particle entropy is
measurable, in the internal state realization of our model, 
by applying a fast Rabi pulse and measuring the amplitude of the ensuing
Rabi oscillations, which is proportional to the Bloch vector 
length $|\s|$.  (Successive measurements with Rabi rotations 
about different axes, i.e. by two resonant pulses differing by a phase 
of $\pi/2$, will control for the dependence on the angle of $\s$).  
In a double well 
realization, one could determine the single-particle entropy by 
lowering the potential barrier, at a moment when the populations 
on each side were predicted to be equal, to let the two parts 
of the condensate interfere. The fringe visibility would then be 
proportional to $|\s|$. Since $|\s|$ at a fixed time depends 
exponentially on $\eta$, such experiments could also potentially
be used to measure scattering lengths.
\begin{figure}
\begin{center}
\epsfig{file=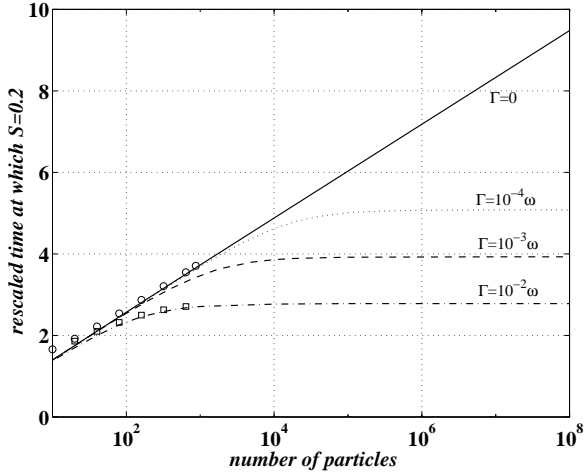,width=0.9\columnwidth}
\end{center}
\caption{ Time at which $S$ reaches 0.2 as a function of the particle 
number $N$, according to the BBR equations (\ref{bgb}), modified to 
include thermal phase-diffusion. Four different 
values of $\Gamma$ are shown: $\Gamma=0$ (-----), 
$\Gamma=10^{-4}\omega$ ($\cdot\cdot\cdot$), $\Gamma=10^{-3}\omega$ ($---$), 
and $\Gamma=10^{-2}\omega$ ($-\cdot-$). Exact quantum results are 
presented for $\Gamma=0$ (circles) and $\Gamma=10^{-2}\omega$ (squares). 
Initial conditions, $\kappa$ and $\omega$ are the same as in 
Figs. 2 and 3.}
\label{f5 }
\end{figure}
Decoherence of quantum systems coupled to reservoirs shows similar behaviour to 
Fig.~(3).  The entropy of a dynamically unstable quantum system coupled to a 
reservoir\cite{PazZurek}, or of a stable system coupled to a dynamically 
unstable reservoir, is predicted to grow linearly with time, at a rate 
independent of the system-reservoir coupling, after an onset time proportional 
to the logarithm of the coupling.  This shows that one can really consider the 
Bogoliubov fluctuations as a reservoir \cite{habib}, coupled to the mean field
with a strength proportional to $1/N$.  But one can also consider decoherence 
due to a genuine reservoir (unobserved degrees of freedom, as opposed to 
unobserved higher moments).  For example, thermal particles scattering off the 
condensate mean field will cause phase diffusion\cite{Anglin} at a rate $\Gamma$ 
which may be estimated in quantum kinetic theory as proportional to the thermal 
cloud temperature.  For internal states not entangled with the condensate 
spatial state, $\Gamma$ may be as low as $10^{-5}$ Hz under the coldest 
experimental conditions, whereas for a double well the rate may reach $10^{-1}$ 
Hz.  Further sources of decoherence may be described phenomenologically with a 
larger $\Gamma$. Evolving the full $N$-particle density matrix under the
appropriate quantum kinetic master equation \cite{Janne}, we again solve for
$\s$ either numerically or in BBR approximation.
In Fig.~4 we show the time at which the entropy reaches a given value, 
as a function of the number of particles, for various $\Gamma$, according 
to the modified BBR equations. The exact quantum results 
(limited by computation power to $N\sim 10^3$) are presented for the 
two limiting values of $\Gamma$, showing excellent agreement with the 
BBR predictions. These results can of course be interpreted as a quantum 
saturation 
of the dephasing rate at low temperature.

This work was partially supported by the National Science Foundation 
through a grant for the Institute for Theoretical Atomic and Molecular 
Physics at Harvard University and Smithsonian Astrophysical Observatory.

\end{document}